\documentclass{article}
\font\bigbf=cmbx10 scaled \magstep3

\usepackage{epsfig}
\usepackage{amssymb}

\begin{document}
\begin{large}
\pagenumbering{arabic}

\rightline{February 19, 2003}
\rightline{to be published in Yad. Fiz.}

\vspace{2.0cm}

\begin{center}

{\large\bigbf {\LARGE{\sl B}}-meson production in $p\bar p$ collisions at Tevatron\\
with $k_T$-factorization}\\

\vspace{1cm}
{\large S.P.~Baranov\footnote{E-mail: baranov@sci.lebedev.ru}

{\it P.N.~Lebedev Institute of Physics,\\
117924 Moscow, Russia\/}\\[3mm]}

\vskip 0.3cm

{\large A.V.~Lipatov\footnote{E-mail: lipatov@theory.sinp.msu.ru}

{\it Physical Department, M.V. Lomonosov Moscow State University,\\
119992 Moscow, Russia\/}\\[3mm]}

\vskip 0.3cm

N.P.~Zotov\footnote{E-mail: zotov@theory.sinp.msu.ru}

{\it D.V.~Skobeltsyn Institute of Nuclear Physics,\\
M.V. Lomonosov Moscow State University,
\\119992 Moscow, Russia\/}\\[3mm]

\end{center}

\vspace{1.0cm}
\begin{center}
{\bf{Abstract}}
\end{center}

In the framework of the $k_T$-factorization QCD approach
we consider the production of $b$ quark pairs in $p\bar p$
collisions at the Fermilab Tevatron.
We investigate the dependence of the $b$ quark, $B$ meson and
decay muon differential cross sections on the different forms of
unintegrated gluon distributions. The analysis also covers the
azimuthal correlations between the $b$ and $\bar b$ quarks and
their decay muons.
Our theoretical results agree well with recent data taken by
the D$\oslash$ and CDF collaborations at Tevatron. Finally, we
present our predictions for muon-muon and muon-jet cross sections
at the Tevatron and CERN LHC conditions.

\vspace{1.5cm}

\section{Introduction} \indent

Recently D$\oslash$ and CDF collaborations have reported new
experimental data~[1--5] on the $b$-flavor production at the Tevatron.
These data are found to be about a factor of two or more larger than
the predictions of perturbation theory (pQCD) at next-to-leading order
(NLO)~[1--6]. Therefore, it would be certainly reasonable to try
a different way.

At the energies of modern colliders (such as Tevatron and LHC), heavy
quark and quarkonium production processes belong to the class of the
so called semihard processes~[7--10]. In these processes, by the
definition, the hard scattering scale $\mu \sim m_Q$ is much larger
than the QCD parameter $\Lambda_{{\rm QCD}}$, but, on the other hand,
it is much smaller than the total center-of-mass energy:
$\Lambda_{{\rm QCD}}\ll \mu\ll\sqrt s$. The last condition
means that these processes occur in the small $x$ region,
$x\simeq m_Q/\sqrt s\ll 1$, where the contributions of the
"large logarithms" to the evolution of gluon densities are known to
become rather important.
It is known also that in the small $x$ region it becomes necessary
to take into account the dependence of the subprocess cross sections
and gluon structure functions on the gluon transverse momentum
$k_T$~[7--10].
Therefore, the $k_T$-factorization (or semihard) approach provides
a more suitable ground for the calculations than the ordinary parton
model.

The $k_T$-factorization approach is based on the
Balitsky-Fadin-Kuraev-Lipatov (BFKL)~[11] evolution equation for
gluon densities. The resummation of the terms
$\alpha_{S}^n\,\ln^n(\mu^2/\Lambda_{{\rm QCD}}^2)$,
$\alpha_{S}^n\,\ln^n(\mu^2/\Lambda_{{\rm QCD}}^2)\,\ln^n(1/x)$ and
$\alpha_{S}^n\,\ln^n(1/x)$ leads to the so called unintegrated
($q_T$-dependent) gluon distributions $\Phi(x,q_T^2,\mu^2)$
which determine the probability to find a gluon carrying the
longitudinal momentum fraction $x$ and transverse momentum $q_T$ at
the probing scale $\mu^2$. In contrast with the usual parton model,
the unintegrated gluon distributions have to be convoluted with
off-mass-shell matrix elements corresponding to the relevant partonic
subprocesses~[7--10]. In the off-mass-shell matrix elements, the
virtual gluon polarization tensor is taken in the form~[7--10]:
\begin{equation}
L^{\mu\,\nu} = {q_T^{\mu}\,q_T^{\nu}\over q_T^2}.
\end{equation}

The $b$-flavor production at the Tevatron in the $k_T$-factorization
approach was considered earlier in~[8, 12--17]. However, the off-shell
matrix elements of hard partonic subprocess presented in~[8]
contain some misprints. Also, the results obtained in the paper~[16]
do not agree with other results~[12--15].

In our previous paper~[17], we analyzed the $p_T$ distribution of the
produced $b$-quarks (presented in the form of integrated cross
sections).
Our results agree well with D$\oslash$~[5] and CDF~[3] experimental data.
Also, we found that our off-shell matrix elements for partonic
subprocess coincide with the ones presented in Ref.~[9].

Here, we use the $k_T$-factorization approach for a more detailed
analysis of the experimental data~[1--5]. We inspect the dependence of
the $b$ quark, $B$ meson and decay muon cross sections on the different
forms of the unintegrated gluon distributions. The analysis also covers
the azimuthal correlations between the $b$ and $\bar b$ quarks and their
decay muons.
Special attention is paid to the role
of the unintegrated gluon distributions which has been applied earlier in
our previous papers~[17--22].
In addition, we present our predictions for muon-muon and muon-jet cross
sections at the Tevatron and CERN LHC conditions.

\begin{figure}[htb]
\begin{center}  
\epsfig{figure= 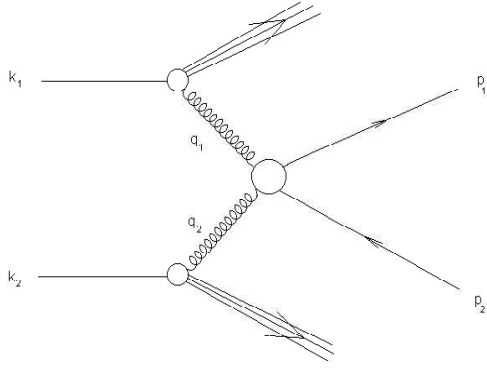,width=8cm,height=6cm}
\end{center}
\caption[]{Diagram for the $p\bar p \to b\bar b\,X$ process.}
\label{eps1}
\end{figure}

The outline of this paper is as follows. In Section 2 we present the
analytical expressions for the total and differential cross sections of
the inclusive heavy quark production in the $k_T$-factorization
approach and describe the unintegrated gluon distributions
which we use in our calculations. In Section 3 we present the
numerical results of our calculations and compare them with the
D$\oslash$~[4, 5] and CDF~[1--3] data. Finally, in Section 4, we
give some conclusions.

\bigskip
\section{Theoretical framework} \indent

In this section we present the expressions for the inclusive heavy quark
production total and differential cross sections in the $k_T$-factorization
approach and describe the unintegrated gluon distributions which we use in
our calculations.

\subsection{Kinematics} \indent

As indicated in Fig.~1, we denote the 4-momenta of the incoming protons
and the outgoing heavy quarks as $k_1$, $k_2$ and $p_1$, $p_2$, respectively.
The initial gluons have the 4-momenta $q_1$ and $q_2$.
We use the Sudakov decomposition, which has the form
\begin{equation}
p_{1} = \alpha_1 k_1 + \beta_1 k_2 + p_{1T},\qquad p_2
= \alpha_2 k_1 + \beta_2 k_2 + p_{2T},\atop\quad {\quad {
q_1 = x_1 k_1 + q_{1T},\qquad q_2 = x_2 k_2 + q_{2T}}},
\end{equation}

\noindent where $p_{1T}$, $p_{2T}$, $q_{1T}$ and $q_{2T}$ are the
transverse (4-vector) momenta of the corresponding particles, and
\begin{equation}
p_1^2 = p_2^2 = m_Q^2,\quad q_1^2 = q_{1T}^2,\quad q_2^2 = q_{2T}^2.
\end{equation}

\noindent In the $p\bar p$ c.m. frame we can write:
\begin{equation}
k_1 = \sqrt s/2\,(1,\,0,\,0,\,1),\qquad k_2 = \sqrt s/2\,(1,\,0,\,0,\,-1),
\end{equation}

\noindent where we neglect the masses of the protons. The Sudakov
variables are expressed as follows:
\begin{equation}
\displaystyle \alpha_1={m_{1T}\over {\sqrt s}}\exp(y_{1}),\qquad
\alpha_2={m_{2T}\over {\sqrt s}}\exp(y_2),\atop
\displaystyle \beta_1={m_{1T}\over {\sqrt s}}\exp(-y_{1}),\qquad
\beta_2={m_{2T}\over {\sqrt s}}\exp(-y_2),
\end{equation}

\noindent where $m_{1,2\,T}^2 = m_{Q}^2 + p_{1,2\,T}^2$, and $y_{1}$ and
$y_2$ are the rapidities of final heavy quarks in the $p\bar p$ c.m. frame.
From the conservation laws, we can easy obtain the following conditions:
\begin{equation}
x_1 = \alpha_1 + \alpha_2,\qquad x_2
= \beta_1 + \beta_2,\qquad  q_{1T} + q_{2T} = p_{1T} + p_{2T}.
\end{equation}

\subsection{Inclusive heavy quark production cross section} \indent

Here we recall some formulas from our previous paper~[17].
In the $k_T$-factorization approach, the differential cross section for
inclusive heavy quark production may be written as
\begin{equation}
\displaystyle d\sigma(p\bar p\to Q\bar Q\,X)
= {1\over 16\pi (x_1\,x_2\,s)^2}\,\Phi(x_1,\, q_{1T}^2,\,\mu^2)\,
\Phi(x_2,\, q_{2T}^2,\,\mu^2)\,\times \atop
\displaystyle \times \sum {|M|^2_{{\rm SHA}}(g^*g^*\to Q\bar Q)}\,
dy_1\,dy_2\,dp_{2T}^2\,dq_{1T}^2\,dq_{2T}^2\,{d\phi_1\over 2\pi}\,
{d\phi_2\over 2\pi}\,{d\phi_Q\over 2\pi},
\end{equation}

\noindent where $\Phi(x_1,q_{1T}^2,\mu^2)$ and
$\Phi(x_2,q_{2T}^2,\mu^2)$ are the unintegrated gluon distributions in
the proton, $\phi_1$, $\phi_2$ and $\phi_Q$ are the azimuthal angles of
the initial gluons and final heavy quark respectively,
$\sum {|M|^2_{{\rm SHA}}(g^*g^*\to Q\bar Q)}$ is the off-mass-shell
matrix element. Symbol $\sum$ in Equ.~(7) indicates the averaging over
the initial and the summation over the final polarization states. The
expression for the $\sum {|M|^2_{{\rm SHA}}(g^*g^*\to Q\bar Q)}$ was
obtained in our previous paper~[17].

Formulas for the differential cross sections in the standard parton model 
(SPM)
may be obtained from Equ.~(7) if we take the limit $q_{1,2\,T}^2\to 0$
and average over the transverse momentum vector $q_{1,2\,T}$:
\begin{equation}
\displaystyle d\sigma(p\bar p\to Q\bar Q\,X)
= {1\over 16\pi\,(x_1 x_2 s)^2}\,x_1G(x_1,\,\mu^2)\,x_2G(x_2,\,\mu^2)\,
\times \atop \displaystyle \times \,
\sum {|M|_{{\rm PM}}^2(gg\to Q\bar Q)}\,dy_1\,dy_2\,dp_{1T}^2\,
{d\phi_Q\over 2\pi},
\end{equation}

\noindent where $\sum {|M|^2_{{\rm PM}}(gg\to Q\bar Q)}$ is the gluon-gluon
fusion matrix element obtained in the standard parton model.

\subsection{Unintegrated gluon distributions} \indent

Various parametrizations of the unintegrated gluon distribution used in
our calculations are discussed below (see also~[23]).

As the first set, we use a BFKL-like parametrization
(hereafter denoted as the JB parametrization) given in~[24].
The method proposed in~[24] lies upon a straightforward perturbative
solution of the BFKL equation where collinear gluon density $xG(x,\mu^2)$
is used as the boundary condition. The unintegrated gluon distribution is
calculated as a convolution of collinear gluon distribution
$xG(x,\mu^2)$ with universal weight factors:
\begin{equation}
\Phi(x,q_T^2,\mu^2) = \int\limits_x^1 \,\varphi(\eta,q_T^2,\mu^2)\,
{x\over \eta}\,G\left({x\over \eta},\,\mu^2\right)\,d\eta,
\end{equation}

\noindent where
\begin{equation}
\displaystyle \varphi(\eta,q_T^2,\mu^2) = \cases{\displaystyle
{\bar \alpha_{S}\over \eta\, q_T^2}J_0
\left(2\sqrt{\mathstrut \bar\alpha_{S}\ln(1/\eta)\ln(\mu^2/q_T^2)} \right),
&if $q_T^2\le \mu^2$,\cr
\displaystyle {\bar\alpha_{S}\over \eta\, q_T^2}I_0
\left(2\sqrt {\mathstrut \bar \alpha_{S}\ln(1/\eta)\ln(q_T^2/\mu^2)}\right),
&if $q_T^2 > \mu^2$,\cr}
\end{equation}

\noindent where $J_0$ and $I_0$ stand for Bessel functions of real and
imaginary arguments respectively, and $\bar\alpha_{S}=3\,\alpha_{S}/\pi$.
In calculations we used the standard GRV set~[25] for $xG(x,\mu^2)$.
The parameter $\bar\alpha_{S}$ is connected with the Pomeron trajectory
intercept: $\Delta = 4\bar\alpha_{S}\ln 2$ in the LO, and
$\Delta = 4\bar\alpha_{S}\ln 2 - N\bar \alpha_{S}^2$ in the NLO
approximations, where $N\sim 18$~[26, 27]. The latter value of $\Delta$
could have dramatic consequences on the high energy phenomenology. However,
some resummation procedures proposed in the last years lead to positive
values: $\Delta \sim 0.2-0.3$~[27, 28]. The result $\Delta = 0.35$ was
obtained from the description of the $p_T$ spectrum of $D^*$ mesons
in the electroproduction at HERA~[29]. We use this value of the
parameter $\Delta$ in the present paper\footnote{We also used this value of
$\Delta$ in the analysis of experimental data on the $J/\psi$ photo- and
leptoproduction at HERA~[19, 20] and in the description~[21, 22] of deep
inelastic structure functions $F_2^c$, $F_L^c$ and $F_L$ in the small $x$
region.}.

Another set of unintegrated gluon densities (the KMS parametrization) [30]
is obtained from a
unified BFKL and DGLAP description of $F_2$ data and includes the so called
consistency constraint [31]. The consistency constraint introduces a large
correction to the LO BFKL equation: about 70\% of the full NLO corrections to
the BFKL exponent $\Delta$ are effectively included in this constraint,
as is explaned in [32].

Finally, the third unintegrated gluon function used here is the one
which is obtained from conventional gluon density $xG(x,\mu^2)$ by taking
the $\mu^2$-derivative~[7, 11, 33]:
\begin{equation}
\Phi(x,q_T^2) = {d{\,xG(x,\mu^2)} \over d{\mu^2}}\bigg\vert _{\mu^2 = q_T^2}.
\end{equation}

\noindent Here we have used the expression for $xG(x,\mu^2)$ from the
standard GRV set~[25]. We point out that the parametrization (11),
in contrast with JB and KMS parametrizations, takes into account
the terms $\alpha_{S}^n\,\ln^n(\mu^2/\Lambda_{{\rm QCD}}^2)$  and
$\alpha_{S}^n\,\ln^n(\mu^2/\Lambda_{{\rm QCD}}^2)\,\ln^n(1/x)$ only.
It is interesting to compare the numerical results obtained with the
parametrizations JB and (11) because they are derived from the same
collinear density but underwent evolution according to different
(BFKL or DGLAP) equations.

The integration limits in (7) and (8) are given by the kinematic
conditions of the D$\oslash$ and CDF experiments~[1--5]. The calculation
of the heavy quark production cross section in the $k_T$-factorization
approach have been done according to (7) for
$q_{1T}^2 \ge Q_0^2$ and $q_{2T}^2 \ge Q_0^2$. For the regions
$q_{1T}^2 \le Q_0^2$ and $q_{2T}^2 \le Q_0^2$, we
set $q_{1T}^2 = 0$ and $q_{2T}^2 = 0$ in the matrix
elements of the hard subprocesses, take
$\sum {|M|_{{\rm PM}}^2(gg\to Q\bar Q)}$
instead of $\sum {|M|_{{\rm SHA}}^2(g^*g^*\to Q\bar Q)}$ and use
equation (8) of the usual parton model. The contributions from the
assymmetric configurations ($q_{1T}^2 \le Q_0^2$, $q_{2T}^2 \ge Q_0^2$
and $q_{1T}^2 \ge Q_0^2$, $q_{2T}^2 \le Q_0^2$) are included
in a similar way, where one of the gluons is described by the
unintegrated distribution and the other one by the collinear density.
The choice of the critical value of the parameter $Q_0^2 = 1\,{\rm GeV}^2$
is determined by the requirement that the value of $\alpha_{S}(\mu^2)$
in the region $q_{1,2\,T}^2 \ge Q_0^2$ be small (where in fact
$\alpha_{S}(\mu^2) < 0.26$).

\bigskip
\section{Numerical results} \indent

In this section we present the numerical results of our calculations
and compare them with the D$\oslash$~[4, 5], CDF~[1--3] and UA1~[34] data.

Besides the choice of the unitegrated gluon distribution, our theoretical
results depend on the bottom quark mass, the factorization scale $\mu^2$ and
the $b$ quark fragmentation function.
For example, a special choice of the $b$-quark fragmentation function was used
in paper~[6] as a way to increase the $B$-meson production cross section
in observable region of transverse momentum. In the present paper we convert
$b$ quarks into $B$-mesons using the usual Peterson fragmentation 
function~[35] with $\epsilon = 0.006$.
Regarding the other parameters, we use $m_b = 4.75\,{\rm GeV}$ and
$\mu^2 = q_{1,2\,T}^2$ as in~[8, 14]\footnote{We also used this choice of
$\mu^2$ earlier~[19, 20] for the description of $J/\psi$ photo- and
leptoproduction processes at HERA within the $k_T$-factorization approach and
the colour singlet model.}.

\begin{figure}[htb]
\begin{center}  
\epsfig{figure= 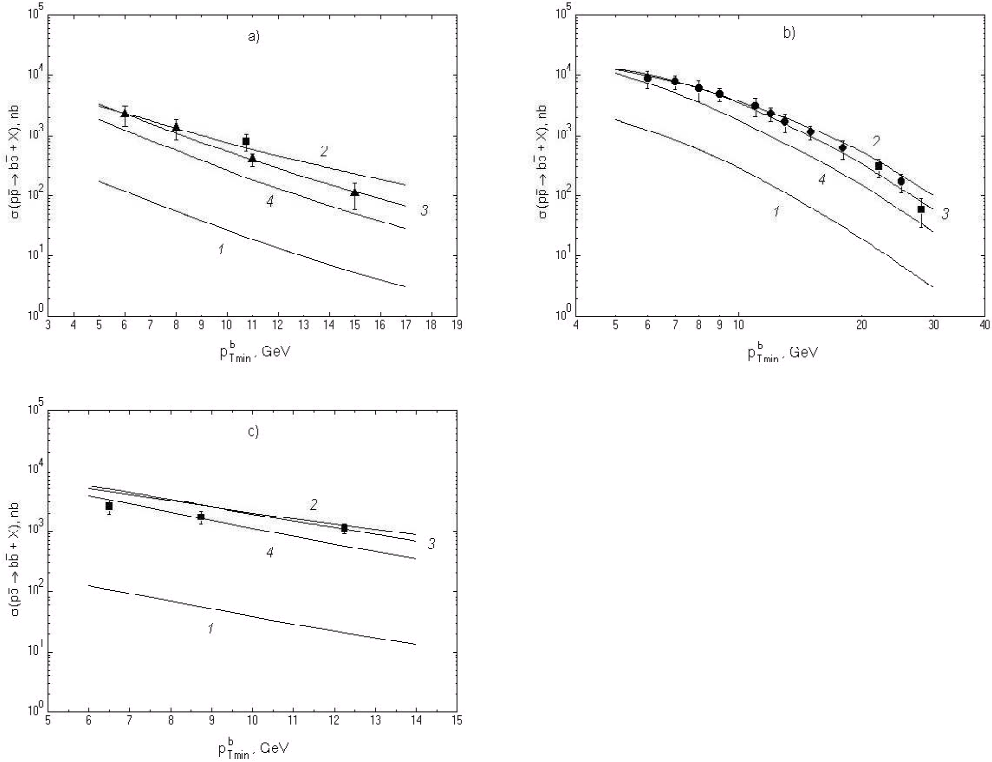,width=14.5cm,height=11cm}
\end{center}
\caption[]{The $b$ quark transverse momentum distributions at
Tevatron conditions presented in the form of integrated cross
sections. The cuts applied: $|y_1| < 1.5$, $|y_2| < 1.5$,
$\sqrt s = 630 \,{\rm GeV}$ (Fig.~2a),
$|y_1| < 1$, $\sqrt s = 1800 \,{\rm GeV}$ (Fig.~2b) and
$|y_1| < 1$, $|y_2| < 1$, $\sqrt s = 1800 \,{\rm GeV}$ (Fig.~2c).
Curve {\sl 1} corresponds to the SPM calculations in the 
leading order approximation with GRV~(LO) gluon density, curves {\sl 2}, 
{\sl 3} and {\sl 4} correspond to the $k_T$-factorization results with JB, KMS
and Equ.~(11) unintegrated gluon distributions. Experimental data are
from UA1~[34] $\blacktriangle$, D$\oslash$~[5] {\LARGE $\bullet$} and CDF~[3] $\blacksquare$.}
\label{eps2}
\end{figure}

The results of our calculations are shown in Fig.~2---9. Fig.~2 displays the
$b$ quark transverse momentum distribution at Tevatron conditions presented in
the form of integrated cross sections. The following cuts were applied:
$|y_1| < 1.5$, $|y_2| < 1.5$,
$\sqrt s = 630 \,{\rm GeV}$ (Fig.~2a),
$|y_1| < 1$, $\sqrt s = 1800 \,{\rm GeV}$ (Fig.~2b) and
$|y_1| < 1$, $|y_2| < 1$, $\sqrt s = 1800 \,{\rm GeV}$ (Fig.~2c).
Curve {\sl 1} corresponds to the SPM calculations at the leading
order approximation with the GRV~(LO) gluon density, curves {\sl 2}, {\sl 3}
and {\sl 4} correspond to the $k_T$-factorization results with the JB, KMS and
the (11) unintegrated gluon distributions, respectively.
One can see that the results obtained in the $k_T$-factorization approach agree
very well with the D$\oslash$ and CDF experimental data. The calculations based
on the parametrization (11), which takes into account only the terms
$\alpha_{S}^n\,\ln^n(\mu^2/\Lambda_{{\rm QCD}}^2)$ and
$\alpha_{S}^n\,\ln^n(\mu^2/\Lambda_{{\rm QCD}}^2)\,\ln^n(1/x)$
predict the cross section which is lower than the data by a factor of about 2
(Fig.~2a and Fig.~2b). We would like to note the difference in the shapes
between the curves obtained using the $k_T$-factorization approach and the
standard parton model. This difference shows the $p_T$ broadening effect which
is usual for the $k_T$-factorization approach. At the same time, the shape of
the curves {\sl 4} and {\sl 1} is practically identical.

\begin{figure}[htb]
\begin{center}  
\epsfig{figure= 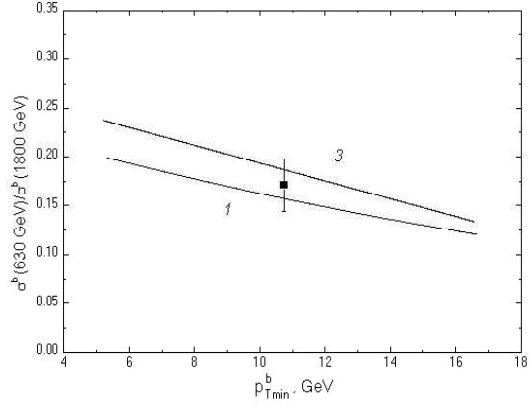,width=8.0cm,height=6.5cm}
\end{center}
\caption[]{The ratio of $\sigma(b)$ at $\sqrt s = 630 \,{\rm GeV}$ to
$\sqrt s = 1800 \,{\rm GeV}$ as a function of the minimum $b$ quark
transverse momentum $p^b_{T\,{\rm min}}$. Notation of the curves {\sl 1} and
{\sl 3} is the same as in Fig.~2. Experimental data are from CDF[1] $\blacksquare$.}
\label{eps3}
\end{figure}

\begin{figure}[htb]
\begin{center}  
\epsfig{figure= 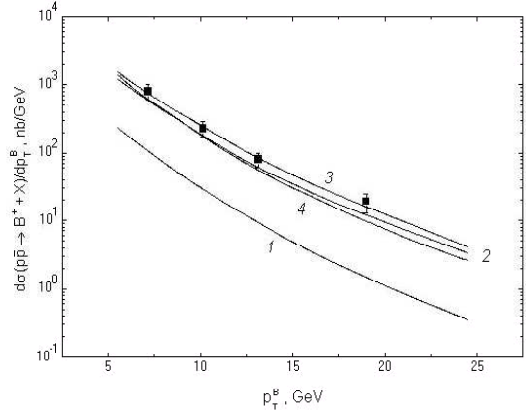,width=8.0cm,height=6.5cm}
\end{center}
\caption[]{Theoretical predictions for the $B$ meson $p_T$ spectrum
at $\sqrt s = 1800 \,{\rm GeV}$ compared to the CDF data. The cuts
applied: $|y^B| < 1$. Notation of the curves {\sl 1 --- 4} is the same as in 
Fig.~2. Experimental data are from CDF~[2] $\blacksquare$.}
\label{eps4}
\end{figure}

It is notable that, in the collinear approximation,
the sum of the LO and NLO pQCD contributions still underestimates the $b$
quark production rate by a factor of 2~[3, 5]. The results obtained in the
$k_T$-factorization approach in Ref.~[16] lie above the sole LO contribution,
but below the sum of LO and NLO contributions, in an apparent disagreement
with our present results. The roots of this discrepancy are connected with the
parameter settings accepted in Ref.~[16], that is, the large values of the
quark mass $m_b = 5\,{\rm GeV}$ and the remormalization scale
$\mu^2 = m_{1,2\,T}^2$ in the running coupling constant.

\begin{figure}[htb]
\begin{center}  
\epsfig{figure= 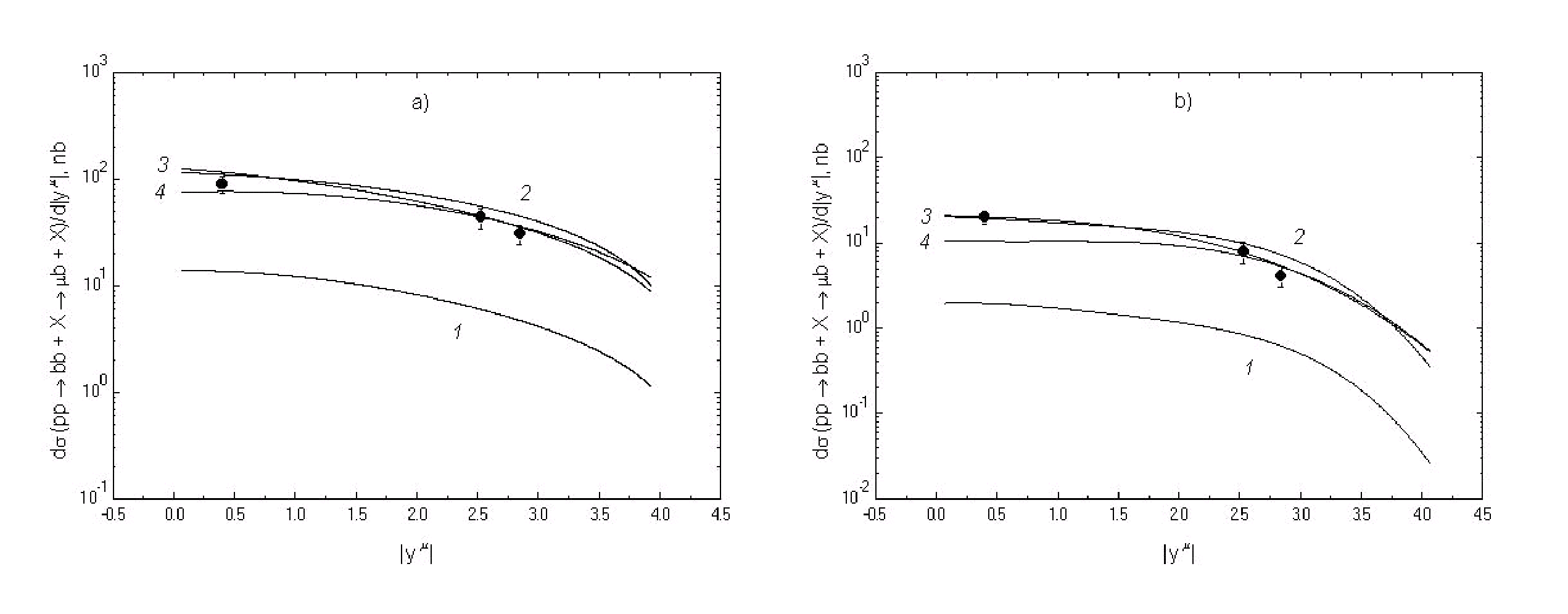,width=14.5cm,height=6.5cm}
\end{center}
\caption[]{The cross section for muons from $B$ meson decay as a function of
rapidity compared to the D$\oslash$ data.
The cuts applied: $p_T^{\mu} > 5 \,{\rm GeV}$ (Fig.~5a) and
$p_T^{\mu} > 8 \,{\rm GeV}$ (Fig.~5b). Notation of the curves {\sl 1 --- 4} 
is the same as in Fig.~2. Experimental data are from D$\oslash$~[4] {\LARGE $\bullet$}.}
\label{eps5}
\end{figure}

Fig.~3 shows the ratio of the cross sections measured at different beam
energies, $\sigma(b)$ at $\sqrt s = 630 \,{\rm GeV}$ to
$\sqrt s = 1800 \,{\rm GeV}$, as a function of the minimum $b$ quark
transverse momentum $p^b_{T\,{\rm min}}$. Notation of the curves {\sl 1, 3}
is the same as in Fig.~2. One can see that the experimental data collected
by the D$\oslash$ collaboration agree with the LO pQCD calculations
as well as with the $k_T$-factorization ones. This result is not surprising
because, when the ratio of the cross sections is considered, many factors
affecting the absolute normaliszation, as well as many theoretical
and experimental uncertainties partially or completely cancel out~[1].

\begin{figure}[htb]
\begin{center}  
\epsfig{figure= 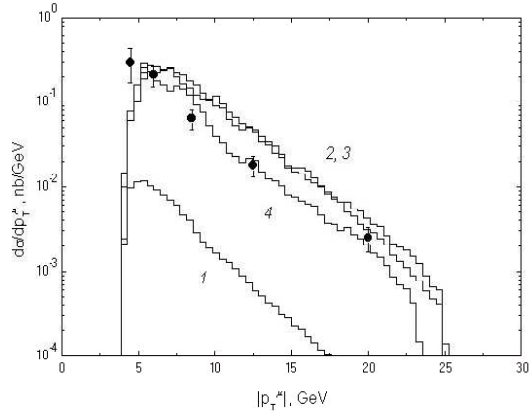,width=8.0cm,height=6.5cm}
\end{center}
\caption[]{Predictions on the leading muon $p_T$ spectrum in the $b\bar b$ production
events compared to the D$\oslash$ data. The cuts applied to both muons:
$4 < p_{T}^{\mu} < 25 \,{\rm GeV}$, $|\eta^{\mu}| < 0.8$ and
$6 < m^{\mu\mu} < 35 \,{\rm GeV}$. Notation of the histograms {\sl 1 --- 4} 
is the same as in Fig.~2. Experimental data are from D$\oslash$~[5] {\LARGE $\bullet$}.}
\label{eps6}
\end{figure}

Fig.~4 shows the prediction for the $B$ meson $p_T$ spectrum
at $\sqrt s = 1800 \,{\rm GeV}$ compared to the CDF data~[3] within the 
experimental cuts $|y^B| < 1$. Notation of the curves {\sl 1 --- 4} is the same 
as in Fig.~2. Here we find good agreement between the results obtained in the
$k_T$-factorization approach and experimental data.
One can see that the $p_T$ broadening effect mentioned earlier appears to be 
not as much clear for the $B$ meson production as it was for the $b$ quark 
production. This is because of compensatory effects in the fragmentation 
process. We note again that the ordinary NLO pQCD calculations 
underestimate the $B$ meson production by a factor of 3~[2].

\begin{figure}[htb]
\begin{center}  
\epsfig{figure= 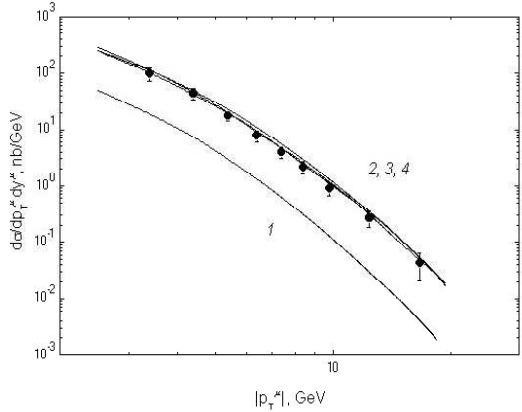,width=8.0cm,height=6.5cm}
\end{center}
\caption[]{Double differential cross section for muons from $B$ meson decay as a function 
of $p_T^{\mu}$ compared to the D$\oslash$ data~[4]. The cuts applied: 
$2.4 < |y^{\mu}| < 3.2$. Notation of the curves {\sl 1 --- 4} is the
same as in Fig.~2. Experimental data are from D$\oslash$~[4] {\LARGE $\bullet$}.}
\label{eps7}
\end{figure}

The recent D$\oslash$ experimental data refer also to muons originating 
from semileptonic decays of $B$-mesons. To produce muons from $B$-mesons in 
theoretical calculations, we simulate their semileptonic decay according to 
the standard electroweak theory. In Fig. 5 we show the rapidity distribution
$d\sigma/d|y^{\mu}|$ of the decay muons for both
$p_T^{\mu} > 5 \,{\rm GeV}$ (Fig.~5a) and
$p_T^{\mu} > 8 \,{\rm GeV}$ (Fig.~5b). 
Notation of the curves {\sl 1 --- 4} is the same as in Fig.~2. We find that 
the $k_T$-factorization results agree very well with the D$\oslash$ 
experimental data. 
In the central rapidity range $|y^{\mu}| < 1$, the results based on 
the parametrization Equ.~(11) lie below the ones based on JB and KMS 
parametrizations by a factor of about 1.5.
The ordinary NLO pQCD calculations lie below the data by a factor of about 
4~[4].

Fig.~6 shows the leading muon $p_T$ spectrum in the $b\bar b$ production
events compared to the D$\oslash$ data, where the leading muon is defined
as the muon with the greatest $p_{T}^{\mu}$ value.
The cuts applied to both muons are $4 < p_{T}^{\mu} < 25 \,{\rm GeV}$, 
$|\eta^{\mu}| < 0.8$ and $6 < m^{\mu\mu} < 35 \,{\rm GeV}$. 
Notation of the histograms {\sl 1 --- 4} is the same as in Fig.~2.
One can see that the histograms {\sl 2} and {\sl 3} lie even a bit higher 
than the experimental data.

\begin{figure}[htb]
\begin{center}  
\epsfig{figure= 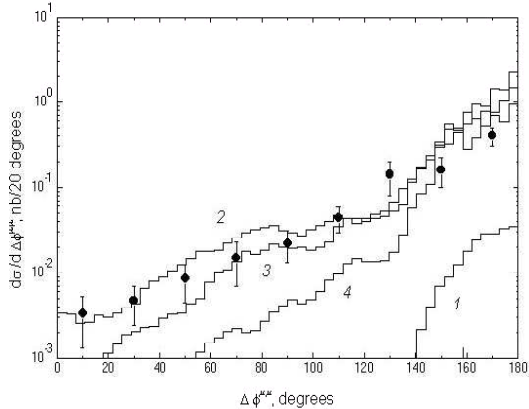,width=8.0cm,height=6.5cm}
\end{center}
\caption[]{Azimuthal muon-muon correlations at Tevatron conditions. 
The cuts applied to both 
muons: $4 < p_{T}^{\mu} < 25 \,{\rm GeV}$,
$|\eta^{\mu}| < 0.8$ and $6 < m^{\mu\mu} < 35 \,{\rm GeV}$.
Notation of the histograms {\sl 1 --- 4} is the same as in Fig.~2.
Experimental data are from D$\oslash$~[4] {\LARGE $\bullet$}.}
\label{eps8}
\end{figure}

The double differential cross sections $d\sigma/dp_{T}^{\mu}\,dy^{\mu}$
in the forward rapidity region $2.4 < |y^{\mu}| < 3.2$ are also
well described by the $k_T$-factorization approach (Fig.~7). Notation of the
curves {\sl 1 --- 4} is the same as in Fig.~2. It is interesting to note that 
the shape of all the curves is practically the same. The NLO pQCD calculations
underestimate the D$\oslash$ data by a factor of 4~[4].

\begin{figure}[htb]
\begin{center}  
\epsfig{figure= 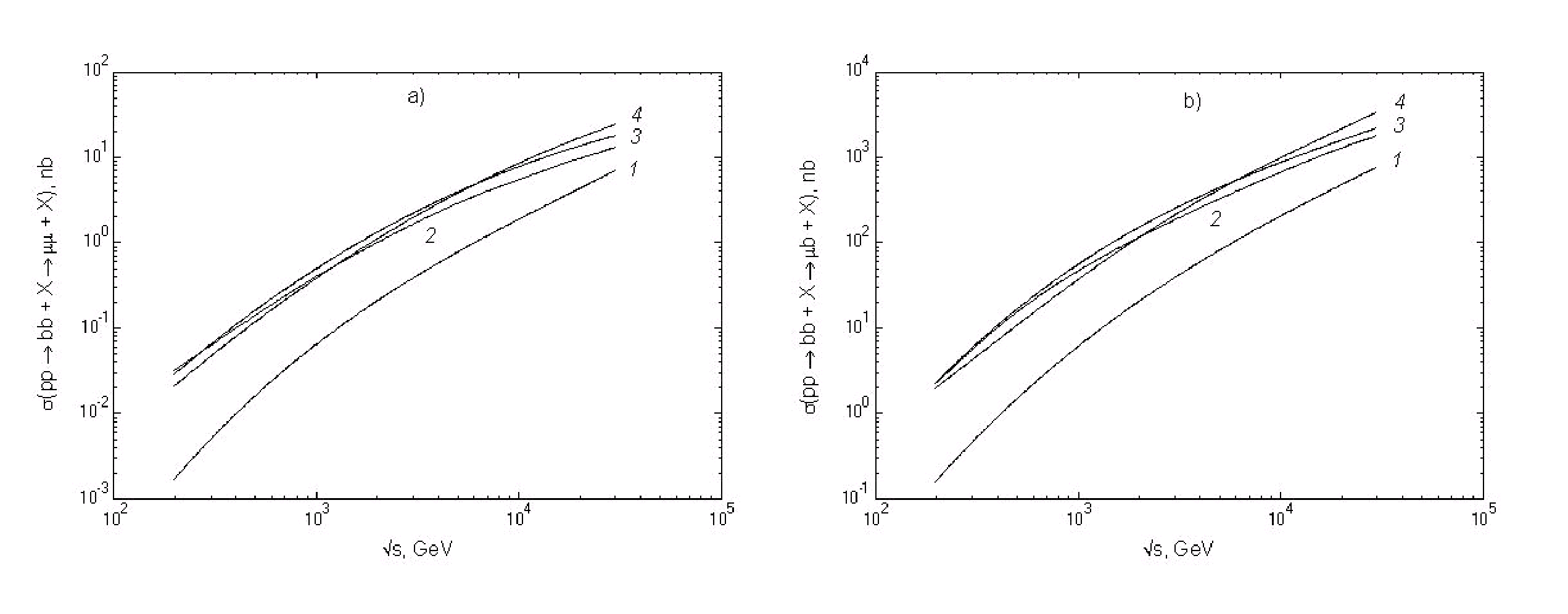,width=14.5cm,height=6.5cm}
\end{center}
\caption[]{Theoretical predictions for muon-muon (Fig.~9a) and muon-jet (Fig.~9b) cross 
sections as functions of $\sqrt s$ at the Fermilab Tevatron and CERN LHC 
conditions. The cuts applied: $p_{T}^{\mu} > 6 \,{\rm GeV}$, $|y^{\mu}| < 2.5$.
Notation of the curves {\sl 1 --- 4} is the same as in Fig.~2.}
\label{eps9}
\end{figure}

We point out that the investigations of $b\bar b$ correlations such as the
azimuthal opening angle between the $b$ and $\bar b$ quarks
(or between their decay muons) allow additional details of the
$b$ quark production to be tested since these quantities are
sensitive to the relative contributions of the different production
mechanisms [8, 12--14, 16].
In the naive gluon-gluon fusion mechanism, the distribution
over the azimuthal angle difference $\Delta \phi^{b\bar b}$ must
be simply a delta function $\delta (\Delta \phi^{b\bar b} - \pi)$.
Taking into account the non-vanishing initial gluon transverse momenta
$q_{1T}$ and $q_{2T}$ leads to the violation of this back-to-back
quark production kinematics in the $k_T$-factorization approach.

The differential $b\bar b$ cross section $d\sigma/d\Delta \phi^{\mu \mu}$
is shown in Fig.~8. The cuts applied to both muons are
$4 < p_{T}^{\mu} < 25 \,{\rm GeV}$, $|\eta^{\mu}| < 0.8$ 
and $6 < m^{\mu\mu} < 35 \,{\rm GeV}$. Notation of the histograms {\sl 1 --- 4}
is the same as in Fig.~2. One can see that good agreement between JB and KMS 
predictions and the experimental data is observed. The shape of the histogram 
{\sl 4} strongly differs from that of the histograms {\sl 2} and {\sl 3}. 
In the small $\Delta \phi^{\mu \mu} \sim 0$ region, the parametrization (11) 
underestimates the D$\oslash$ experimental data.
This fact indicates the importance of the large $\alpha_{S}^n\,\ln^n(1/x)$
contributions. One can see that the properties of different unitegrated gluon 
distributions manifest themselves in the $b\bar b$ or muon-muon azimuthal 
correlations. As expected, the sole LO pQCD contribution predicts a peak at
$\Delta \phi^{\mu \mu} \sim \pi$.

In addition, we present our predictions for muon-muon (Fig.~9a) and
muon-jet (Fig.~9b) cross sections at the Tevatron and CERN LHC conditions. 
In the latter case, we take the kinematic requirements
of the detector ATLAS~[36] as a representative example.
This conditions imply the presence of a decay muon with
$p_{T}^{\mu} > 6 \,{\rm GeV}$ and $|y^{\mu}| < 2.5$.
Curves {\sl 1 --- 4} are the same as in Fig.~2.
We point out that the predicted cross sections are rather uncertain and
approximate because we neglected the saturation effects in the gluon 
distributions~[8, 12, 13]. However, the treshold value of $x$ (where these 
effects will come into play) is still unknown.

\bigskip
\section{Conclusions} \indent

In this paper we have considered heavy quark production in $p\bar p$
collisions at Tevatron in the framework of the $k_T$-factorization approach.
We investigated the dependence of the $b$ quark, $B$ meson and
the decay muon cross sections on different forms of the
unintegrated gluon distribution. The analysis covered the azimuthal
correlations between the $b$ and $\bar b$ quarks and their decay muons.
We compared the theoretical results with recent experimental data
collected by the D$\oslash$ and CDF collaborations at Tevatron. We found that 
the $k_T$-factorization results agree well with the experimental data
when we use JB or KMS unintegrated gluon distributions and set the quark mass
$m_b = 4.75\,{\rm GeV}$, the factorization scale $\mu^2 = q_T^2$ and
$\Lambda_{{\rm QCD}} = 250\,{\rm MeV}$. The properties of different
unitegrated gluon distributions manifest themselves 
in the $b\bar b$ or muon-muon azimuthal correlations.
From the analysis of these correlations we can conclude that
JB and KMS unintegrated gluon distributions are more preferable
than the parametrization (11). Finally, we present our predictions
for the muon-muon and muon-jet cross sections at
Tevatron and CERN LHC conditions.

\bigskip
\section{Acknowledgments} \indent

The study was supported in part by RFBR grant ${\rm N}^{\circ}$
02--02--17513. A.V.L. also was supported by INTAS grant YSF 2002 ${\rm N}^{\circ}$ 399.

\end{large}
\end{document}